\documentclass[11pt,preprint]{aastex}
\usepackage{epsfig}
\usepackage{rotating}      
\tightenlines
\newcommand     \amax   {a_{\rm max}}
\newcommand     \amin   {a_{\rm min}}
\newcommand     \Angstrom       {\,{\rm \AA}}

\newcommand     \beq    {\begin{equation}}
\newcommand     \beqa   {\begin{eqnarray}}
\newcommand     \betadisk {{\beta_{\rm disk}}}
\newcommand     \betaism {{\beta_{\rm ism}}}
\newcommand     \cm     {\,{\rm cm}}

\newcommand     \dist   {{d}}
\newcommand     \eeq    {\end{equation}}
\newcommand     \eeqa   {\end{eqnarray}}

\newcommand     \g              {\,{\rm g}}
\newcommand     \GHz    {{\,\rm GHz}}
\newcommand     \gtsim  {\gtrsim}                %apj version

\newcommand     \K              {\,{\rm K}}
\newcommand     \km     {\,{\rm km}}

\newcommand     \ltsim  {\lesssim}               %apj version
\newcommand     \mm     {\,{\rm mm}}

\newcommand     \nH     {n_{\rm H}}

\newcommand     \s              {\,{\rm s}}

\newlength{\figwidth}
\countdef\decade=200
\advance\decade by \year
\countdef\hours=201
\advance\hours by \time
\divide\hours by 60
\countdef\mins=202
\advance\mins by \hours
\multiply\mins by 60
\multiply\hours by 100
\countdef\miltime=203
\advance\miltime by \hours
\advance\miltime by \time
\advance\miltime by -\mins

\begin{document}

\title{
%------------- enable for labelling preprint ---------------------------
        \vspace*{-3.0em}
        {\normalsize\rm submitted to {\it The Astrophysical Journal}}\\ 
        \vspace*{1.0em}
%-----------------------------------------------------------------------
         On the Submillimeter Opacity of Protoplanetary Disks
%        \\
%        {\small DRAFT: \todayd}
       }

\author{B. T. Draine}
\affil{Princeton University Observatory, Peyton Hall, Princeton,
NJ 08544; {\tt draine@astro.princeton.edu}}

\begin{abstract}
Solid particles with the composition of interstellar dust
and power-law size
distribution $dn/da\propto a^{-p}$ for $a\leq\amax$ 
with $\amax\gtsim3\lambda$ and $3<p<4$
will have submm opacity
spectral index $\beta(\lambda)\equiv d\ln\kappa/d\ln\nu\approx(p-3)\betaism$, 
where $\betaism\approx1.7$
is the opacity spectral index of interstellar dust material in the
Rayleigh limit.
For the power-law index $p\approx3.5$ that characterizes
interstellar dust, and that appears likely for particles growing by
agglomeration in protoplanetary disks,
grain growth
to sizes $a\gtsim 3\mm$ 
will result in $\beta(1\mm)\ltsim 1$.
Grain growth can naturally account for
$\beta\approx1$ observed for protoplanetary disks, provided that
$\amax \gtsim 3\lambda$.
\end{abstract}

\keywords{planetary systems: protoplanetary disks
        dust, extinction,
        submillimeter}

\section{Introduction
        \label{sec:intro}}

From mm to mid-infrared wavelengths, the radiation emitted by protoplanetary
disks
is primarily to due to thermal emission from solid particles, just
as for the interstellar clouds out of which the protostars
form.  If the emitting medium is optically thin, then the observed flux 
density $F_\nu$ is proportional to the mass $M_d$ of emitting particles,
\beq
F_\nu \approx \kappa(\nu) M_d B_\nu (T_d)\dist^{-2} ~~~,
\eeq
where $\kappa(\nu)$ and $T_d$ are the opacity and temperature
of the solid material, $B_\nu$ is the blackbody intensity, 
and $\dist$ is the distance.
At submm wavelengths, the Rayleigh-Jeans limit 
$h\nu\ll kT_d$ usually applies, and
\beq
F_\nu \approx \frac{2k}{c^2} \nu^2 \kappa(\nu) \frac{M_d T_d}{\dist^2}
\eeq
is proportional to the product $M_d T_d$.
If $T_d$ can be estimated
from the observed spectrum or other considerations, 
the observed $F_\nu$ allows
$M_d$ to be determined, provided that $\kappa$ and $\dist$ are known.

If the opacity has a power-law dependence on frequency,
$\kappa(\nu) \propto \nu^\beta$, the observed flux $F_\nu \propto \nu^\alpha$,
with $\alpha=2+\beta$.  Even if the absolute opacity cannot be
determined, multiwavelength observations allow the opacity spectral
index $\beta(\lambda)\equiv d\ln\kappa/d\ln\nu$ to be measured.

Observations of the infrared and submm emission from diffuse interstellar
clouds (with masses estimated independently from 
observations of H~I and CO)
allow the opacity $\kappa$ of interstellar dust to be determined.
The observed far-infrared and submm emission from diffuse clouds
is consistent with $\beta\approx 1.7$ 
(Finkbeiner, Davis, \& Schlegel 1999; Li \& Draine 2001).

The opacity spectral index $\beta$ can also be determined for dust
in dense molecular clouds.
The dust
near embedded infrared sources in the Orion ridge has
$\beta\approx 1.9$ between 1100$\micron$ and $450\micron$ 
(Goldsmith, Bergin \& Lis 1997).
The dust in and around 17 ultracompact H~II regions 
studied by Hunter (1998) has
$\beta\approx 2.00\pm 0.25$ between 350 and 450$\micron$.
Friesen et al.\ (2005) find $\beta=1.6_{-0.3}^{+0.5}$ for
the dust in 3 hot molecular cores, with densities $\nH\approx10^8\cm^{-3}$.
The dust in both diffuse clouds and dark clouds thus
appears to be consistent
with an opacity spectral index $\betaism\approx 1.8\pm0.2$ in the submm.

If the dust in protoplanetary disks were both optically thin and
similar to interstellar dust, 
we would expect
$\alpha=2+\betaism\approx3.8\pm0.2$.
However, Beckwith and Sargent (1991) found that protoplanetary disks
usually have spectra with $2<\alpha<3$.
The opacity spectral index $\betadisk$ was estimated by fitting simple disk
models; the 24 objects studied had a median $\betadisk = 0.92$, with
2/3 of the sample falling in the interval $\betadisk=0.9\pm0.7$.

Numerous subsequent studies have also found $\betadisk < \betaism$.
% btd 05.07.12 change post-ApJ-submission
% Natta \& Testi (2004) find a median $\betadisk=0.7$ for a sample of 9 disks 
Natta et al.\ (2004) find a median $\betadisk=0.7$ for a sample of 9 disks
around Herbig Ae stars
%----------------------------------------
observed with the VLA.
Submm interferometry of a rotating disk in the star-forming region
IRAS 18089-1732 indicates $\alpha-2\approx 3.2$ in the outer envelope,
with small-scale structure having $\alpha-2\approx0.5$ (Beuther et al.\ 2004).
Andrews \& Williams (2005)
obtained submm
SEDs for 44 protoplanetary disks in Taurus-Auriga, 
and found that they can be fit using disk models 
with $\betadisk\approx1$.

There are several possible explanations for the 
difference between $\betaism$ and $\betadisk$:
\begin{enumerate}
\item Some of the emission may come from regions that are not optically
thin, even at mm wavelengths.
This appears to be part of the explanation,
but 
radiative transfer models also require material
with $\betadisk \approx 1$ in order to reproduce the observed spectra 
(e.g., Beckwith \& Sargent 1991; 
%btd 05.07.12 change post-ApJ-submission
Natta et al.\ 2004;
%---------------------------------------
Natta \& Testi 2004; 
Andrews \& Williams 2005).

\item
The chemical 
composition of the particles in the disk might be very different from
the composition of interstellar grains.

\item Growth by coagulation in protoplanetary disks 
might lead to ``fluffy'' structures, with
the difference between $\betaism$ and $\betadisk$ 
being due to differing grain geometry.

\item As has been proposed by a number of authors
(e.g., Beckwith \& Sargent 1991; Miyake \& Nakagawa 1993;
D'Alessio et al.\ 2001;
%btd 05.07.12 change post-ApJ-submission
Natta et al.\ 2004;
%---------------------------------------
Natta \& Testi 2004)
grain growth might have put an appreciable fraction of the
solid mass into particles that are so large
that they are not in the Rayleigh limit, even at mm wavelengths,
leading to reduced $\beta$.

\end{enumerate}

Here we show that 
power-law size
distributions $dn/da\propto a^{-p}$ for $a\leq \amax$,
with $3<p<4$ and $\amax\gtsim3\lambda$, will lead to
opacity
spectral index $\beta\approx(p-3)\beta_s$, where $\beta_s$
is the opacity spectral index of the solid material in the
Rayleigh limit.
Therefore, starting with material with the observed $\beta_s\approx 1.7$ of
interstellar dust, grain growth with the likely power-law index
$p\approx3.5$ will produce a mixture of sizes with
$\beta\approx0.9$ if $\amax\gtsim3\lambda$.
Since grain growth is expected to be rapid in protoplanetary disks
(e.g., Beckwith et al.\ 2000; Dullemond \& Dominik 2005),
this can naturally explain
$\beta\approx1$ observed for protoplanetary disks, without need to
appeal to changes in composition or grain geometry, other than size.

We briefly review what is expected regarding the 
optical constants of candidate materials at submm wavelengths
in \S\ref{sec:betas}, and conclude that changes in composition are
not likely to acccount for the low $\beta$ values in disks.
Particle size distributions
are briefly reviewed in \S\ref{sec:size distributions}, to motivate the
adopted power-law distribution.  
In \S\ref{sec:analytic estimate},
the opacity is estimated 
analytically for power-law
size distributions, to find the expected behavior.
In \S\ref{sec:numerical examples} we calculate the opacity numerically
for several candidate materials.
Although the dependence on size and wavelength can
be complicated for conducting materials (in particular, amorphous carbon),
we can reproduce the observed $\betadisk\approx 1.0$ for 
candidate materials with very different optical constants.
Our results are discussed in \S\ref{sec:discussion} 
and summarized in \S\ref{sec:summary}.

\section{\label{sec:betas}
         Dielectric Functions: What is Expected for $\beta_s$?}

Let $\beta_s(\lambda)\equiv d\ln\kappa/d\ln\nu$ in the small-particle limit,
where $\kappa$ is the absorption cross section per unit mass of solid
material.
Simple models of insulators and simple models of conductors both have
$\beta_s=2$ at low frequencies (Draine 2004).  How do real materials
behave at submm wavelengths?

If the low values of $\beta$ observed in protoplanetary disks are to
be produced by materials with small values of $\beta_s$, this would
require materials with enhanced low-frequency absorption.
In insulators, this would require optically-active low energy modes.
Amorphous solids, with low-temperature specific heats revealing a spectrum
of low-energy excitations (Pohl \& Salinger 1976), are candidates for
enhanced low-frequency absorption.

\begin{figure}[h]
\begin{center}
\epsfig{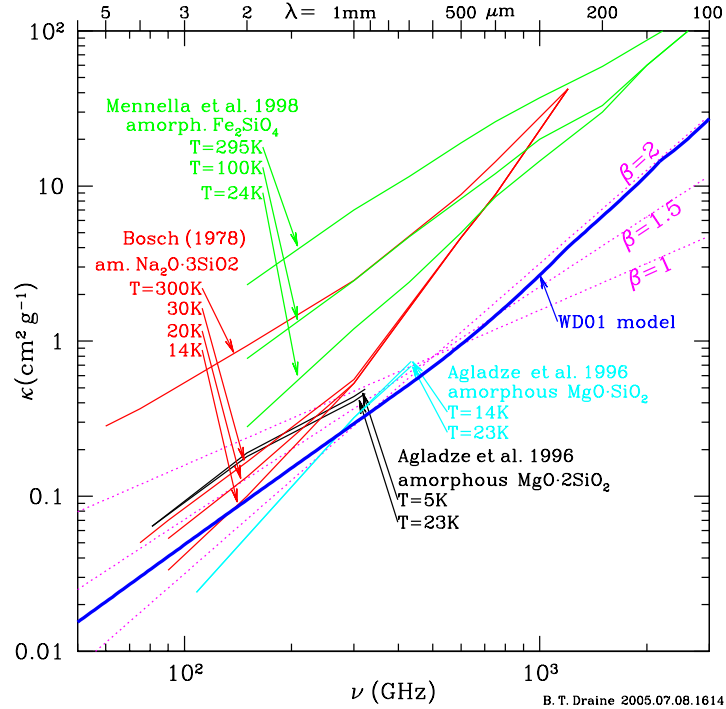}
\caption{\label{fig:kappalabins}
         \footnotesize
         Laboratory measurements at various temperatures 
	 of opacities of amorphous Fe$_2$SiO$_4$
         (Mennella et al.\ 1998), Na$_2$O$\cdot$3SiO$_2$ (B\"osch 1978),
         and MgO$\cdot$2SiO$_2$ and MgO$\cdot$SiO$_2$ (Agladze et al.\ 1996).
         Curve labelled WD01 is the opacity of the 
	 Weingartner \& Draine (2001) dust mixture, for comparison.
         Dotted lines show power laws with $\beta=1$, 1.5, and 2.}
\end{center}
\vspace*{-2.em}
\end{figure}

Laboratory measurements of the opacity of amorphous insulators are
shown in Figure \ref{fig:kappalabins}.  In some cases (e.g.,
MgO$\cdot$2SiO$_2$; Agladze et al.\ 1996) 
the submm opacity is consistent with a power law over the wavelength range 
studied.
In other cases, broken power laws provide an acceptable fit.
However, in all cases the power law index $\beta_s>1$.
The smallest effective value is $\beta_s\approx 1.2$ for
MgO$\cdot$2SiO$_2$ between 150 and 300 GHz (Agladze et al.\ 1996), but
for most samples, $\beta_s\approx 2$, with $\beta_s>2$ seen
in some cases (e.g., Na$_2$O$\cdot$3SiO$_2$ between 300 and 1200 GHz:
B\"osch 1978).
While the number of materials studied 
is limited, it appears that small particles of
amorphous insulators 
are likely to have $\beta_s \gtsim 1.5$ in the submm.

\begin{figure}[h]
\begin{center}
\epsfig{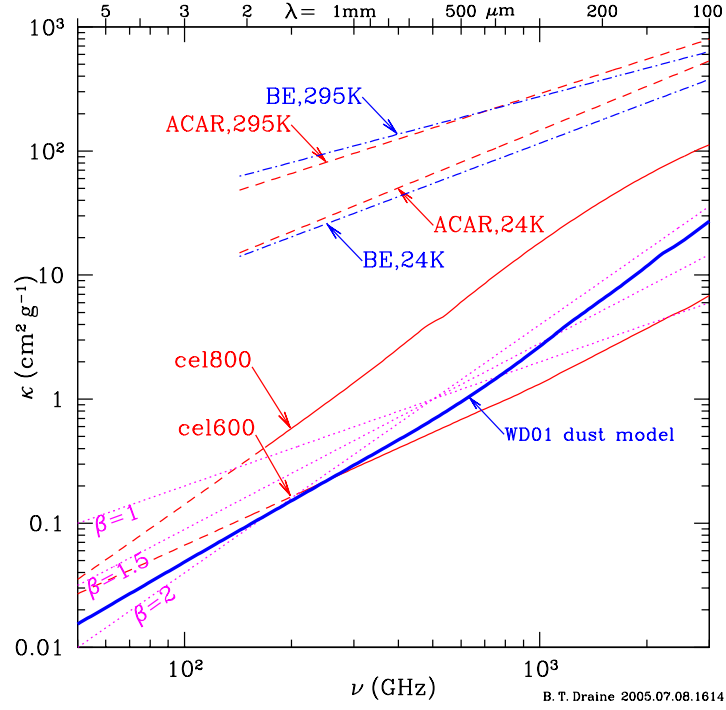}
\caption{\label{fig:kappalabcarb}
         \footnotesize
         Same as Figure \ref{fig:kappalabins} but
         for carbonaceous materials BE and ACAR measured at
	 $T=24\K$ and $T=295\K$
	 (Mennella et al.\ 1998), and calculated for small spheres
         with optical constants measured for
	 cellulose pyrolyzed at 600C and 800C 
	 (J\"ager et al.\ 1999) -- see text.
        }
\end{center}
\vspace*{-2.em}
\end{figure}

For carbonaceous materials the situation is less clear.  Opacities
measured by Mennella et al.\ (1998) for materials ``BE'' 
(soot produced by burning C$_6$H$_6$ in air) and
``ACAR'' (amorphous carbon grains produced from a carbon arc in Ar gas)
are shown in Figure \ref{fig:kappalabcarb}.
The opacities are very large (e.g., at 1mm, $\kappa\approx 25-30\cm^2\g^{-1}$
for BE and ACAR at 24K, a factor 20--100 larger than 
for the insulators in Fig.\ \ref{fig:kappalabins} at $T<30\K$).
Samples BE and ACAR are characterized by small $\beta_s$: at 24K,
BE and ACAR have $\beta_s\approx 1.1$ and 1.2, respectively.
There is reason for concern that the very large submm opacities seen
in the lab for materials BE and ACAR may be the result of contact
between the particles in the sample (which could dramatically affect
the opacity for materials with $|\epsilon|\gg 1$) in which case the
measured opacities do not correspond to those that would be produced
by isolated small spheres of the same material.

J\"ager et al.\ (1999) have produced amorphous carbonaceous solids with
varying levels of graphitization by pyrolysis of cellulose, and measured
the optical constants.
Pyrolysis at 600C in Ar gas produces a fine-grained material ``cel600''
with C:H:O::1:0.33:0.053 and density $\rho=1.67\g\cm^{-3}$.
Although the material is substantially ``graphitized'' (in the sense that
$\sim70\%$ of the C atoms appear to be $sp^2$ bonded, as in graphite),
the material has $\epsilon_2 \rightarrow 0$ as $\nu\rightarrow0$, implying
zero d.c. conductivity.
J\"ager et al.\ provide the refractive index for 
$0.2\micron<\lambda<450\micron$.  
We extrapolate to lower frequencies assuming
$\epsilon_1(\lambda \geq 450\micron)\approx 4.03$ and
$\epsilon_2(\lambda \geq 450\micron)=  0.0378(450\micron/\lambda)^{0.3}$,
giving $\beta_s=1.3$ for $\lambda > 450\micron$.
The opacity calculated for small spheres of this material is shown in
Figure \ref{fig:kappalabcarb}.
As noted by J\"ager et al., small spheres have $\beta_s \approx 1.5$ for
$\lambda \gtsim 100\micron$.

Pyrolysis at 800C produces a material ``cel800'' with C:H:O::1:0.14:0.030,
with $\rho=1.84\g\cm^{-3}$ (J\"ager et al.\ 1999).
The estimated 
$sp^2$ fraction is $\sim75\%$, and the material
becomes conducting.  J\"ager et al.\ measured the optical constants for
$0.178\micron <\lambda < 562\micron$; 
we extrapolate to lower frequencies with
a simple free-electron model with conductivity 
$\sigma=3.6\times10^{13}\s^{-1}$,
and damping time $\tau=1.5\times10^{-13}\s$.
The opacity calculated for small spheres of cel800 is shown
in Figure \ref{fig:kappalabcarb}.
At 1\,mm, this calculated opacity is more than an
order of magnitude below what was reported for ACAR and BE.
As noted by J\"ager et al., $\beta_s=1.95$ for $\lambda > 100\micron$.

With $|\epsilon|\gg 1$ in the submm, $\kappa$ for cel800
is strongly shape-dependent.  Because of its simple analytic properties, the
``Continuous Distribution of Ellipsoids'' (CDE) is sometimes used to
allow for nonspherical shapes. J\"ager et al.\ show that for the CDE,
cel800 would have $\beta_s \approx 0.7$, but
caution that the CDE serves only as an illustrative
example of shape effects.
The CDE assumes that a small fraction of the ellipsoids have
very extreme shapes, and needle-like particles
make a large contribution to $\kappa$ when $|\epsilon|\gg 1$
(Min, Hovenier, \& de Koter 2003).
The shape distribution of interstellar particles does not seem
likely to resemble the CDE, at least as regards these extreme
shapes. We consider that spheres (or ellipsoids with modest axial ratios)
provide a better approximation to interstellar particles than the CDE.

Any material capable of explaining $\beta\ltsim 1$ on the basis of
composition rather than size
would have to be unlike the insulators in Figure 1, or the pyrolyzed cellulose
in Figure 2.
The opacities reported by Menella et al.\ for carbonaceous materials
BE and ACAR are large and have $\beta\approx 1$, but it is not clear
that the measured opacity would apply to free particles and, if it did,
such material could not be a significant component of interstellar dust.
Although one cannot exclude the possibility that 
protoplanetary disks might include material
with both large mm/submm opacity and $\beta_s\approx 1$, alternate
explanations should be considered.

\section{\label{sec:size distributions}
         Size Distribution}

Consider particles with a power-law size distribution,
\beq \label{eq:dn/da}
\frac{dn}{da} = A \left(\frac{a}{\amax}\right)^{-p}
~~~ \amin \leq a \leq \amax ~~~,
\eeq
where $A$ is a constant; the size distribution presumably results
from competition between growth (e.g., coagulation) and destruction
(e.g., shattering).  

Experimental studies of fragmentation find power-law size distributions for
the fragments,
with $p$ varying between $\sim 1.9$ for low-velocity collisions, to
$p\approx 4$ for catastrophic impacts (Davis \& Ryan 1990).
However, the power-law index 
for fragments from
shattering events integrated over a range of target masses need not
be the same as
the power-law index for fragments of a single target.

Dohnanyi (1969) showed that, for certain assumptions, coagulation and
collisional fragmentation would lead to
a steady-state size distribution with $p=3.5$. 
Tanaka et al.\ (1996) argue
that the $p=3.5$ power law is in fact a very general result that
depends {\it only} on the assumption that the fragmentation
process is self-similar,\footnote{
  If the collision rate varies as $a^q$, then
  $p=2.5+q/2$ (Tanaka et al.\ 1996).
  }
and that the collision rate varies as
$a^2$.  Self-similarity may 
not apply to larger
bodies, where self-gravity becomes important, or to the smallest sizes,
where surface free energy considerations will limit the numbers of
very small fragments, but self-similarity may apply, at least
approximately, to the cratering and shattering events
occuring in the interstellar
medium and in protoplanetary disks.

Extinction and scattering of starlight by interstellar dust can be
reproduced
by a mixture of graphite and silicate grains 
with $p\approx3.5$ and $\amax\approx0.25\micron$
(Mathis, Rumpl, \& Nordsieck 1977; Draine \& Lee 1984).
Although more recent work 
(e.g., Kim \& Martin 1995; 
Weingartner \& Draine 2001)
find deviations from a simple power law with cutoffs,
the $p=3.5$ power law gives a good accounting for the overall
size distribution spanning many orders of magnitude in mass.
While the origin of the interstellar 
grain size distribution remains uncertain, it is
intriguingly
close to the predictions of self-similar fragmentation.

The distribution of $D>30\km$ asteroids has 
$p\approx 3.4$ (Dohnanyi 1969).
Based on observations by the Sloan Digital Sky Survey (Ivezi\'c et al.\ 2001),
the asteroid size distribution 
appears to be consistent with 
$p\approx 3.25$ for 
diameters $D$ between $5$ and 
$300\km$
(Bottke et al.\ 2005).
%
% btd note: if extend to smaller sizes, the observed distribution seems
% appears to get less steep:
% $p\approx 2.9$ over the range 1-50$\km$ 
%

If the size distribution of particles in a protoplanetary disk is
determined by a competition between coagulation and fragmentation,
it is plausible that the size distribution might be close to a power
law with $p\approx 3.5$, with an upper cutoff $\amax$ that will depend on
how far coagulation has proceeded.
Note that for $p\approx 3.5$, most of the mass is at the large size
end of the distribution, with the numbers of large grains nearly
independent of the lower cutoff $\amin$, provided
$\amin \ll\amax$.

\section{\label{sec:analytic estimate}
         Analytic Approximation for the Opacity}

For a size distribution $dn/da$, the opacity (absorption cross section
per unit mass of solid materials) at frequency $\nu$ is given by
\beq \label{eq:kappa}
\kappa(\nu) \equiv
\frac{\int da (dn/da) C_{\rm abs}(a,\nu)}
     {\int da (dn/da) (4\pi/3)\rho a^3} ~~~,
\eeq
where $\rho$ is the solid density, and $a\equiv (3m/4\pi\rho)^{1/3}$ is the
effective radius of a 
particle of mass $m$.

Consider some frequency $\nu$, wavelength $\lambda=c/\nu$, and
suppose that the absorption cross section for particles of size $a$
can be 
approximated by
\beq \label{eq:Cabs}
C_{\rm abs}(a,\nu) \approx 
\left\{
\begin{array}{ll}
  (4\pi/3) a^3 \alpha_\nu & {\rm for~}
 a < a_c(\nu)
\\
\pi a^2 &{\rm for~} a > a_c(\nu)
\end{array}
\right.
\eeq
where 
\beq
\alpha_\nu \equiv 
\frac{18\pi}{\lambda}\frac{\epsilon_2}{(\epsilon_1+2)^2+\epsilon_2^2}
~~~{\rm and}~~~
a_c \equiv \frac{3}{4\alpha_\nu} \equiv \frac{\lambda}{24\pi}
\frac{(\epsilon_1+2)^2+\epsilon_2^2}{\epsilon_2}
\eeq
are defined so that eq.\ (\ref{eq:Cabs}) is
exact for spheres in the small particle limit ($a/\lambda\rightarrow 0$)
(see, e.g., Draine \& Lee 1984).
Here $\epsilon_1(\lambda)$ and $\epsilon_2(\lambda)$ 
are the real and imaginary parts of
the complex dielectric function 
$\epsilon\equiv \epsilon_1 + i\epsilon_2$.
Eq.\ (\ref{eq:Cabs}) is only approximate for
$a\gtsim \lambda/2\pi$, but the limiting behavior in the geometric optics
limit $a\rightarrow \infty$
is sensible, corresponding to a low-albedo object.
Because we will be interested in size distributions where the surface
area is dominated by small particles, our analytic estimate of the
opacity integral (\ref{eq:kappa}) is relatively insensitive to errors in 
the albedo of the particles with $a\gg \lambda/2\pi$.
For $a\approx \lambda/2\pi$, $C_{\rm abs}$ can exceed
eq.\ (\ref{eq:Cabs}) by as much as a factor $\sim10$ (depending on
$\epsilon$), but for the moment we use
eq.\ (\ref{eq:Cabs}) to estimate the
opacity for the size distribution (\ref{eq:dn/da}):
\beq
\kappa(\nu) = \frac{\alpha_\nu}{\rho} ~~~{\rm if}~~\amax < a_c
\eeq
and
\beq
\kappa(\nu) =
\frac{\alpha_\nu}{\rho}
\frac{1}{(p-3)}
\left(\frac{a_c}{\amax}\right)^{4-p}
\left[1 - 
      (p-3)\left(\frac{\amin}{a_c}\right)^{4-p} -
      (4-p)\left(\frac{\amax}{a_c}\right)^{3-p}
\right]
\eeq
for $\amin < a_c < \amax$, $p>3$, $p\neq 4$, and $\amin\ll\amax$.
If $3 < p < 4$, 
then
\beq
\kappa(\nu) \approx 
\frac{3}{4\rho \amax} \frac{1}{(p-3)}
\left(\frac{4\alpha_\nu \amax}{3}\right)^{p-3}
~~~{\rm if}~~
\label{eq:amax condition}
\amin \ll a_c \ll \amax ~~~.
\eeq
If the small-particle absorption cross section per
volume varies as a power-law in frequency,
\beq
\alpha_\nu = \alpha_0 \left(\frac{\nu}{\nu_0}\right)^{\beta_s}
~~~,
\eeq
then
\beq \label{eq:result}
\kappa(\nu)  = 
\frac{\alpha_0}{(p-3)\rho} 
\left(\frac{3}{4\alpha_0\amax}\right)^{4-p}
\left(\frac{\nu}{\nu_0}\right)^{(p-3)\beta_s}
~~~{\rm if}~~
\amin \ll a_c \ll \amax ~~~.
\eeq
Therefore, if $\amin \ll a_c \ll \amax$, the size distribution will have
\beq \label{eq:beta=(p-3)betas}
\beta\approx (p-3)\beta_s
~~~.
\eeq
For $p=3.5$ and $\beta_s\approx 1.7$,
eq.\ (\ref{eq:beta=(p-3)betas}) predicts $\beta\approx 0.85$:
material with the FIR-submm 
opacity characteristic of diffuse interstellar
dust ($\beta_s \approx 1.7$) can reproduce the observed frequency dependence of
the opacity of material in circumstellar disks ($\betadisk \approx 1$)
{\it if}
$\amax$ is large enough so
that $a_{\max}\gg a_c(\lambda)$ over the frequency range of interest.

The derivation of eq.\ (\ref{eq:result}-\ref{eq:beta=(p-3)betas}) relied on the approximations
(\ref{eq:Cabs}), which are known to not be quantitatively accurate.
We now proceed to direct calculation of $\kappa(\nu)$ for several candidate
materials to determine the actual 
frequency dependence of $\beta$,
and to see what values of $\kappa$ result when $\amax$ is large
enough to satisfy the condition (\ref{eq:amax condition}).

\section{\label{sec:numerical examples}
         Examples}

The precise behavior of $\kappa(\nu)$ depends on the
complex dielectric function $\epsilon(\nu)$.
Not knowing the precise composition of interstellar and
protostellar grain
material, we consider three, quite different, candidate materials.  
We will see that for the different materials considered here, 
a power-law size distribution with
$p\approx 3.5$ results in mm/submm opacities with $\beta\approx 0.9$
provided $\amax \gtsim 3\mm$.

The opacity $\kappa(\nu)$ (eq.\ \ref{eq:kappa}) is evaluated with
$C_{\rm abs}(a,\nu)$ calculated for spheres, using
Mie theory for $a < a_x\equiv 2\times10^4\lambda/2\pi |m|$, 
and geometric optics for $a>a_x$,
where
$m=\sqrt{\epsilon}$ is the complex refractive index.
We employ a double precision version of 
the Mie theory implementation of Wiscombe (1980, 1996).

\subsection{Silicate Particles}

\begin{figure}[h]
\begin{center}
\epsfig{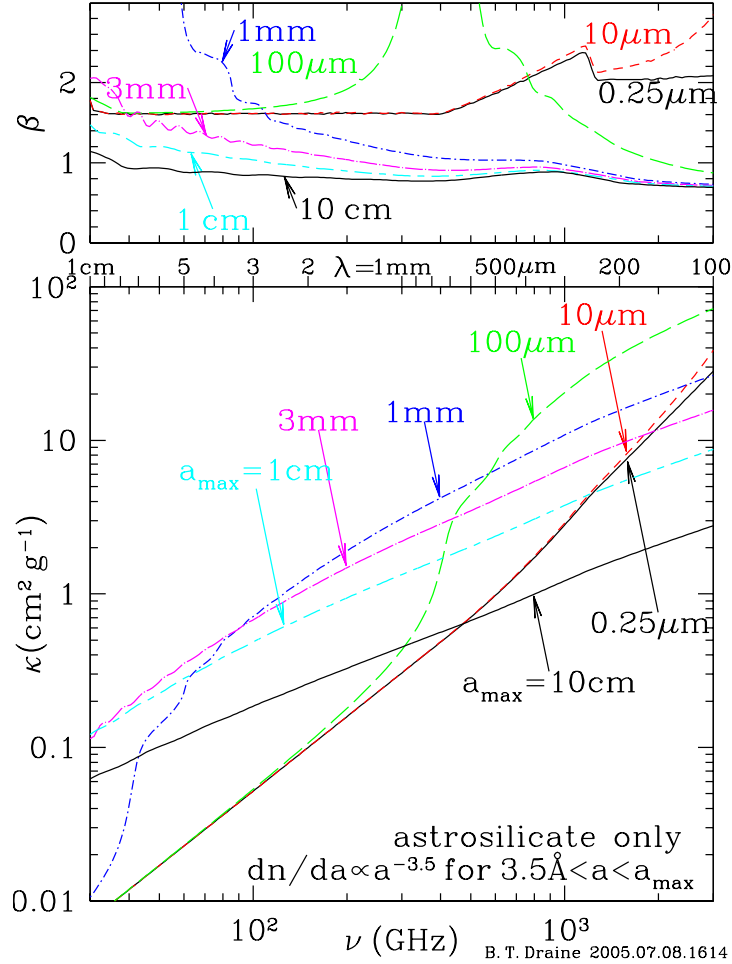}
\caption{\label{fig:astrosil}
         \footnotesize
         Opacity of amorphous silicate spheres with size distribution
         $dn/da \propto a^{-3.5}$ for $3.5\Angstrom <a < \amax$.
         Curves are labelled by $\amax$;
	 curve for $0.25\micron$ is the small-particle limit,
	 and numerically close to the WD01 dust model
	 (not shown).
	 Upper panel shows $\beta\equiv d\ln\kappa/d\ln\nu$ for
	 selected $\amax$.
	 $\beta(1\mm) \ltsim 1$ is found for $\amax \gtsim 3\mm$.
	 }
\end{center}
\end{figure}

We calculate the opacity of spherical silicate particles
using the dielectric function estimated for interstellar silicate
material 
(Draine 2003).
Figure \ref{fig:astrosil} shows $\kappa(\nu)$ for size distributions with
$p=3.5$, $\amin=3.5\Angstrom$, and various values of $\amax$.
The curve labelled $0.25\micron$ is the ``small particle'' 
opacity estimated for interstellar
silicate grains.  At $\lambda= 1\mm$, 
increasing $\amax$ does not noticeably change
the opacity until $\amax$ reaches $100\micron$, at which point the
optics of small particles with $2\pi a/\lambda \approx 1$ results in an
{\it increase} in the opacity.  This effect leads to opacities for the overall
size distribution that {\it exceed}
the small-particle opacity at 1\,mm until $\amax$ reaches $10\cm$.
For $\amax \gtsim 1\mm$, $\kappa(1\mm)\propto 1/\sqrt{\amax}$,
consistent with eq.\ (\ref{eq:result}).
The upper panel in Fig.\ \ref{fig:astrosil} shows the opacity
spectral index $\beta(\lambda)$ for the different choices of $\amax$.
For this material, the condition $a_c\ll\amax$ can be replaced by
$3\lambda\ltsim\amax$.
We see that $\beta(1\mm) \ltsim 1$ for $\amax\gtsim3\mm$.

\begin{figure}[h]
\begin{center}
\epsfig{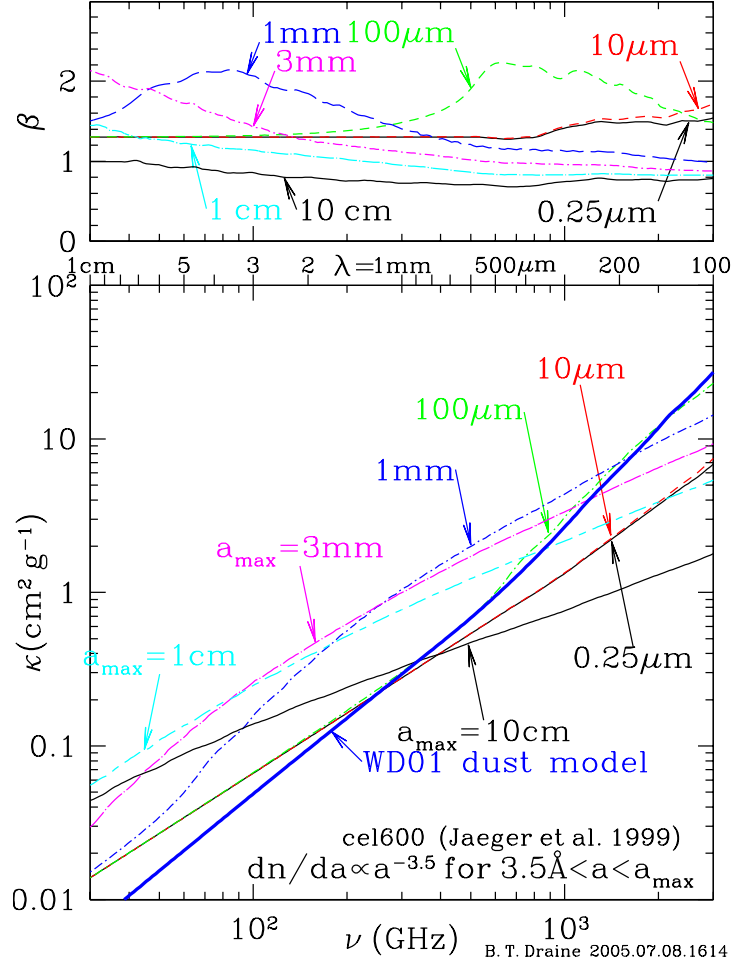}
\caption{\label{fig:pcel600}
         \footnotesize
	 Same as Fig.\ \ref{fig:astrosil}, but using
         the optical constants of
         cellulose pyrolyzed at 600C (J\"ager et al.\ 1999).
	 Also shown is the opacity for the WD01 dust model.
	 For $\amax \leq 100\micron$, $\kappa(1\mm)$
	 is very similar to the WD01 opacity.
         For $\amax \gtsim 3\mm$, the opacity has
	 $\beta(1\mm) \ltsim 1$.}
\end{center}
\end{figure}
\subsection{Amorphous Carbonaceous Material}

The dust model of Weingartner \& Draine (2001; hereafter WD01) includes
carbonaceous material with
the anisotropic dielectric function of graphite, with 
absorption and scattering
by randomly-oriented spherical grains estimated using the 
``1/3-2/3 approximation'' (Draine \&
Malhotra 1993).  Graphite is a conductor, with appreciable conductivity in
the basal plane, and nonzero electrical conductivity
even parallel to the c axis (i.e., normal to the basal plane).
For large particles at submm wavelengths, the conductivity plays a
major role in determining the absorption cross section.
Because large particles formed by coagulation will not have the properties of
crystalline graphite, we consider instead the amorphous carbonaceous
solids cel600 and cel800 
studied by J\"ager et al.\ (1999)
(see \S\ref{sec:betas}).

Figure \ref{fig:pcel600} shows the FIR-submm opacity for size distributions
of cel600 spheres.
In the small-particle limit,
the opacity of cel600
near $\lambda=1\mm$ is similar to or less than the
opacity of the WD01 grain model, and at 100$\micron$ the opacity
is less than the WD01 model by a factor $\sim3$.
The WD01 model successfully 
reproduces the observed emission from the ISM
(Li \& Draine 2001) and other spiral galaxies (Regan et al.\ 2004).
Candidate interstellar grain materials that do not
substantially exceed 
the WD01 opacity could, in principle, be abundant in the ISM.
Thus, the mm/submm properties of interstellar carbonaceous material 
could perhaps resemble cel600.  
The mm/submm opacity of the size distributions of cel600
is insensitive to $\amax$ until $\amax$ reaches
$\sim100\micron$.
Size distributions of cel600 with
$p=3.5$ have $\beta(1\mm)\ltsim1$
for $\amax \gtsim 1\cm$.

\begin{figure}[h]
\begin{center}
\epsfig{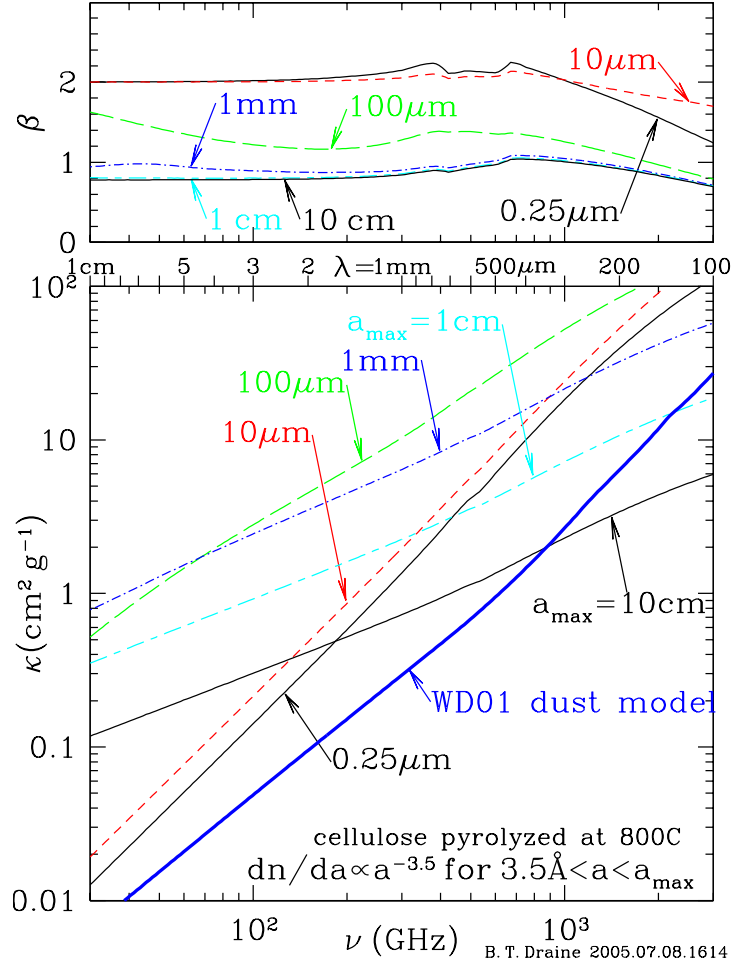}
\caption{\label{fig:pcel800}
        \footnotesize
        Same as Fig.\ \ref{fig:pcel600}, but for cellulose pyrolyzed
        at 800C (J\"ager et al.\ 1999), 
	resulting in a material with appreciable d.c.
	conductivity.
	In this case, magnetic dipole absorption (see text)
	causes $\kappa(\lambda=1\mm)$ 
	to increase as $\amax$ is increased beyond $1\micron$.
	These particles have $\beta(1\mm)<1.2$ for $\amax \gtsim 200\micron$.
	}
\vspace*{-1.em}
\end{center}
\end{figure}
\begin{figure}[h]
\begin{center}
\epsfig{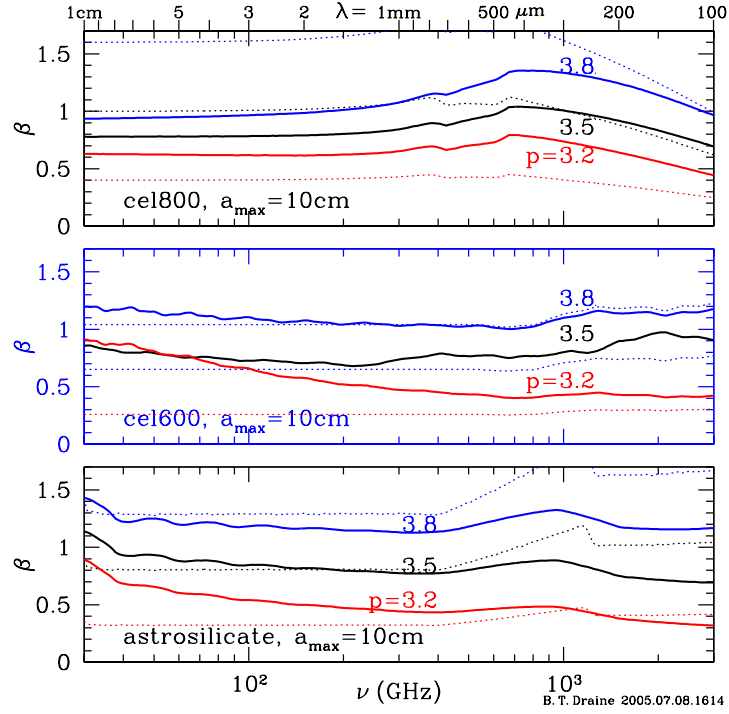}
\caption{\label{fig:beta for different p}
        \footnotesize
	$\beta(\nu)$ for size distributions $dn/da\propto a^{-p}$,
	for 3 values of $p$, and the materials of Figs.\
	\ref{fig:astrosil}--\ref{fig:pcel800}.
	Dotted lines are $(p-3)\beta_s$, which should be approximately
	equal to $\beta(\lambda)$ (see eq.\ \ref{eq:beta=(p-3)betas}), 
	for
	$p=3.2,3.5,3.8$ (bottom to top in each panel) showing
	approximate agreement with the actual $\beta$ values.
	}
\vspace*{-1.em}
\end{center}
\end{figure}

Spheres composed of cel800, on the other hand, gives FIR and submm
absorption that is much stronger than in the WD01 dust model, even when the
material is in the form of spheres -- any other shape would provide
even stronger absorption.
For $1\mm\ltsim\lambda\ltsim100\micron$, in the small-particle limit
$\kappa$ exceeds the WD01 opacity by factors of $\sim4$; hence 
it does not seem likely that the interstellar grain mix could include 
very much such material.
Nevertheless, because we cannot rule out the possiblity that
protoplanetary disk material might have optical properties
resembling cel800, we consider size distributions of spheres
composed of cel800 as an example of a conducting material.
In the small-particle limit, cel800 has $\beta_s\approx 2$,
but size distributions with
$\amax\gtsim 3\mm$ give $\beta(1\mm) \approx0.9$, showing the generality of the
result (\ref{eq:beta=(p-3)betas}).

To see the sensitivity of $\beta$ to the power-law index $p$,
Fig.\ \ref{fig:beta for different p} shows $\beta$ for $p=3.2$, 3.5, and 3.8.
The opacity spectral index $\beta(1\mm)<1$ requires $p\ltsim 3.6$ for
astrosilicate or cel600, or $p\ltsim3.8$ for cel800.

The relation $\beta\approx(p-3)\beta_s$ was derived using 
eq. (\ref{eq:Cabs}), which is not an accurate approximation for $C_{\rm abs}$
when $a\gtsim\lambda/2\pi$.
Also shown in Fig.\ \ref{fig:beta for different p} (dotted lines)
are $(p-3)\beta_s(\nu)$ for $p=3.2$, 3.5, and 3.8.
We see that while $(p-3)\beta_s$ is often close to the actual $\beta(\nu)$,
there are regimes where
$\beta$ differs appreciably from this estimate 
(e.g., astrosilicate with $p=3.8$ at $\nu > 10^3\GHz$;
cel600 with $p=3.2$ for $\nu<10^2\GHz$, 
and cel800 with $p=3.8$ for $\nu<500\GHz$).
Nevertheless eq.\ (\ref{eq:beta=(p-3)betas}) provides a reasonable
estimate.

\section{\label{sec:discussion}
         Discussion}
\subsection{Ices?}
In diffuse clouds, H$_2$O and other ices are not detected: 
ultraviolet radiation prevents the accumulation of H$_2$O ice coatings
on grains in diffuse clouds.
In dark clouds, however, substantial amounts of solids containing
H$_2$O, CO, CH$_3$OH, H$_2$CO, NH$_3$, CH$_4$, and other compounds
are observed to be present, presumably as ``mantles'' coating 
``cores'' (silicate and carbonaceous material) inherited from
diffuse clouds.

What will be the role of these ices at submm wavelengths?
Using the dielectric function of
crystalline H$_2$O ice~I, Aannestad (1975)
found that the presence of ice mantles 
had only minor effects on the absorption at $\lambda \gtsim 200\micron$,
because the vibrational and hindered-rotation modes of the H$_2$O
all lie at higher frequencies.
This is confirmed in recent work (Pontoppidan et al.\ 2005).

Preibisch et al.\ (1993) considered silicate grains coated with
``dirty ice'' consisting of a mixture of H$_2$O and NH$_3$ with
small amorphous carbon inclusions.  The optical properties of this
mixture were estimated using effective medium theory.
In the submm, the absorption by this
``dirty ice'' material was almost entirely due to the amorphous carbon
inclusions.  When present as inclusions within an ice matrix, 
the amorphous carbon material is more absorptive than when
present as a bare particle.  It should be kept in mind, however, that
this conclusion is based on use of approximate effective medium theories to 
estimate the optical properties of the ice with inclusions.

Therefore, although 
the absorption cross section for a silicate or carbonaceous grain
will change if an ice mantle develops on the grain, at submm
frequencies it does not appear that this will substantially alter the
grain absorption cross section unless the ice itself is strongly
absorbing, and strong absorption does not appear likely unless the ``ice'' is
contaminated by amorphous carbon particles -- in which case the
absorption is due to the amorphous carbon, not the ice itself.

\subsection{Grain Growth}

Beckwith \& Sargent (1991) suggested that grain growth might explain
the values of $\betadisk\approx 1$ inferred from their observations.
Miyaka \& Nakagawa (1993) calculated the opacity for size distributions
extending to $\amax\gtsim 1\cm$, and concluded that grain growth
could indeed account for the observed values of $\betadisk$.
Miyaka \& Nakagawa considered grains composed of mixtures of silicate and
H$_2$O ice, either as compact solids or in fluffy structures.
The present paper shows that the result of Miyaka \& Nakagawa is quite
general, and applies to both conducting and nonconducting materials.

D'Alessio et al.\ (2001)
calculated the opacity for power-law size distributions with $p=2.5$ and
$p=3.5$, for a mixture of separate spheres of silicate, ice, and organic 
material, but for the silicate material they 
assumed constant $\epsilon_2$ for $\lambda>800\micron$, resulting in
$\beta_s=1$ for the silicate spheres at $\lambda > 800\micron$.
Therefore even in the small particle limit, their model had
$\beta_s=1$ for $\lambda>800\micron$.  
This frequency dependence is not expected -- see
Fig.\ \ref{fig:kappalabins} where amorphous insulators have
$\beta_s\approx 2$ at mm wavelengths.

%btd 05.07.12 change post-ApJ-submission
%Natta \& Testi (2004) 
Natta et al.\ (2004)
%---------------------------------------
calculated $\kappa$ and $\beta$ for 
power-law size distributions with various 
exponents $p$ for mixtures of compact spheres of olivine, organic
material, and water ice, and also for fluffy grains containing mixtures of
the above, with 50\% vacuum.
They used dielectric functions such
that $\beta_s=1.27$ for the mixture of separate, compact 
%btd 05.07.12 change post-ApJ-submission
%spheres.
spheres (Natta \& Testi 2004).
%---------------------------------------
They showed that by increasing $\amax$ to 
$\amax\gtsim 3\mm$ for the compact grains, or $\amax\gtsim 10\cm$
for the fluffy grains, size distributions with $p\approx3.5$ would have
$\beta(1\mm)\approx1$.

The results of the present paper are in general agreement with these
previous studies.
We also point out that when $\amax\gtsim\lambda$,
the approximate estimate $\beta\approx(p-3)\beta_s$ applies
to both insulating and conducting particles (for $3<p<4$).
For the likely power-law index $p\approx3.5$,
this gives $\beta\approx0.5\beta_s$,
so that materials with $\beta_s=2$, for example, are
expected to have $\beta\approx 1$ when $\amax\gg\lambda$.
Direct calculation of $\kappa$ for size distributions of three
candidate materials -- silicate, nonconducting carbonaceous material
(cel600),
and conducting carbonaceous material (cel800) confirm that
power-law size distributions with $p\approx3.5$ will have
$\beta\ltsim1$ for $\amax\gtsim3\lambda$.

It is sometimes suggested that the low values of $\betadisk$ are due
to ``fluffiness'' of grains grown by agglomeration (Beckwith \&
Sargent 1992).
Fluffiness could subtantially increase the mm/submm opacity
if the grain material is electrically conducting
(Wright 1987), but for nonconducting materials with 
dielectric functions that are not very
large at the wavelengths of interest, grain geometry will have only
modest effects on the grain opacity.  
Kr\"ugel \& Siebenmorgen (1994)
considered extinction and absorption by ``fluffy'' grains composed of
silicate, amorphous carbon, and ice.  
The absorptivity of the ice in
the far infrared was treated as a free parameter, and chosen to be
large enough for the ice absorption to dominate the submm absorption.
With their adopted optical constants, the particles had 
$\beta_s\approx 1.8$ for
$\lambda>100\micron$.  
They found that power-law size distributions
(\ref{eq:dn/da}) with $p=3.5$ and $\amax = 3\mm$ had $\beta\approx0.5$
for $100\micron \ltsim \lambda \ltsim 2\mm$.

Stognienko et al.\ (1995) discussed the optical properties of particles 
produced by coagulation of smaller sub-particles, and showed that the
opacity of such clusters depends strongly on the optical properties
of the sub-particles, and on the geometry of the clustering.
For amorphous carbon sub-particles, the particle geometry can lead to
strong enhancements in absorption for some cluster topologies.
However, if the cluster size $a\ll\lambda$, 
the clusters appear to have $\beta\approx 2$
at submm frequencies.
Henning \& Stognienko (1996) calculated the opacity for dust aggregates;
their standard model has $\beta\approx 2$ for the aggregates.

From the above studies we conclude that, while grain structure
unquestionably plays a role in determining scattering and
absorption cross sections, the observed small values
of $\betadisk$ are naturally explained simply as the result of increased
grain size, independent of whether the grains are compact or
conglomerate.
Grain growth to sizes $\gtsim 3\mm$ can naturally account
for the observed $\betadisk(1\mm)\approx 1$.

%It should also be kept in mind that the grains in the diffuse interstellar
%medium may also include products of coagulation, but are observed to have
%$\betaism\approx 1.7$.

\section{\label{sec:summary}Summary}

The principal conclusion of this paper is the following: 
if the solid particles
in protostellar disks have size distributions $dn/da \propto a^{-p}$ for
$a<\amax$, with $p\approx 3.5$ (as observed for interstellar dust, and
as seems likely for particles in protoplanetary disks)
and $\amax\gtsim 3\mm$, the resulting
opacities have 
$\beta(1\mm)\approx (p-3)\beta_s\approx 1$,
where $\beta_s$ is the opacity spectral index in the small particle limit.
Therefore if the particles in protoplanetary disks are composed of the
same material as interstellar grains (with $\beta_s \approx 1.8\pm0.2$
in the submm) and the size distribution has $p\approx3.5$ and
extends to $\amax\gtsim3\mm$, 
the protoplanetary disk material will have
$\beta(1\mm)\ltsim 1$.
This result is not expected to be altered by the addition of
of dielectric ice coatings on the grains.

To demonstrate the generality of this conclusion, three quite different
materials were considered: two amorphous insulators (astrosilicate and
cellulose pyrolyzed at 600C) 
and a conductor (cellulose pyrolyzed
at 800C).
While the relation $\beta\approx(p-3)\beta_s$ is only an approximation,
we find that all 3 materials give $\beta(1\mm)\ltsim1$ for $p\approx3.5$
and $\amax\gtsim3\mm$.

The observed submm emission from
protostellar disks therefore is explained naturally
by the particle growth that
is expected to occur in these disks.

\acknowledgements
This research was supported in part by NSF grant AST-0406883.
I am grateful to 
%btd 05.07.12 change post-ApJ-submission
A. Natta for helpful discussions, and to
%---------------------------------------
R.H. Lupton for availability of the SM software package.

\end{document}